\documentstyle[twocolumn,aps,prl,psfig,epsf,epsfig]{revtex}
\draft
\begin{document}
\title{Bose Condensates with 1/r Interatomic Attraction:
Electromagnetically Induced ``Gravity''}
\author{
D. O'Dell$^{(1)}$ \and S. Giovanazzi$^{(1)}$ 
\and G. Kurizki$^{(1)}$ \and V. M. Akulin$^{(2)}$ }
\address{
(1) Department of Chemical Physics, Weizmann Institute of Science,
76100 Rehovot, Israel\\
(2) Laboratoire Aim\'{e} Cotton, CNRS II, Orsay Cedex 91405, France}
\date{\today}
\maketitle
\begin{abstract}
We show that particular configurations of intense off-resonant
laser beams can give rise to
an attractive $1/r$ interatomic potential  between atoms located
well within the laser wavelength.
Such a ``gravitational-like'' interaction is shown to give  stable
Bose condensates that are \emph{self-bound} (without an additional
trap) with unique scaling properties and measurably distinct 
signatures.
\end{abstract}

In the atomic Bose-Einstein condensates (BECs) 
created thus far \cite{anderson95} the atoms interact only at 
very short distance in good correspondence with the hard sphere model.
The majority of the properties of these dilute gases 
can be understood by taking into account only two-body collisions 
which are characterized by the s-wave scattering length \cite{dalfovo99}.
A number of groups \cite{tiesinga92} have investigated the fascinating
possibility of
changing the magnitude and sign of the s-wave scattering length
using external fields. The resulting condensates retain
the essential hard-sphere, s-wave, nature
of the interatomic interaction.

Here we wish to introduce a qualitatively new regime of cold atoms
in which the atom-atom interactions are attractive and have a very long range,
varying as $r^{-1}$ \cite{goral99}.

We shall demonstrate that a stable BEC with attractive $r^{-1}$
interactions
is achievable by irradiating the atoms with
intense, \emph{off-resonant}, electromagnetic fields.
The atoms are then coupled
via the dipoles that are induced by these external fields
(in contrast to those induced by the random vacuum field
responsible for the van der Waals-London interaction,
which varies as $r^{-6}$, and leads to the usual
hard sphere description) \cite{thirunamachandran80}.

Such an $r^{-1}$ attractive potential can simulate gravity
between quantum particles. Remarkably, this potential gives an
interatomic attraction
(depending on the laser intensity and wavelength) which
can be some 17 orders of magnitude greater than 
their gravitational interaction at the same distance.

This suggests it might be possible
to study gravitational effects, normally only important on the
stellar scale, in the laboratory. 
Particularly interesting is the possibility of experimentally
 emulating Boson stars 
\cite{ruffini69}: gravitationally bound condensed Boson configurations of 
finite volume, in which the zero-point kinetic energy balances the 
gravitational attraction and thus \emph{stabilizes} the system against 
collapse.

In this letter we shall discuss the interplay of the usual hard-core 
interatomic potential with an electromagnetically induced 
``gravitational'' one on a BEC using a variational 
mean-field approximation (MFA).
Two new physical regimes with unique scaling properties emerge where the BEC
is self-bound (no trap required).

How can one realize the $r^{-1}$ potential between neutral atoms? 
Consider the dipole-dipole interaction energy induced by the presence
of external electromagnetic radiation of intensity $I$, wavevector
${\bf q}$, and polarization $\hat{{\bf e}}$. This energy can be written
(in S.I. units) in terms of cartesian components $i,j$
\cite{thirunamachandran80}
\begin{equation}
U({\bf r}) =
\left(\frac{ I}{4 \pi c \varepsilon_{0}^{2}}\right)
\alpha^{2}(q) \hat{{\bf e}}_{i}^{\ast} \hat{{\bf e}}_{j} V_{ij}(q, {\bf r})
\cos ({\bf q} \cdot {\bf r}).   \label{eq:tpot}
\end{equation}
Here ${\bf r}$ is the interatomic axis, $\alpha(q)$ the isotropic,
dynamic, polarizability of the atoms at frequency $cq$, and
$V_{ij}$ is the retarded dipole-dipole interaction tensor
\begin{eqnarray}
V_{ij}=\frac{1}{r^{3}} & \Big[ \big(\delta_{ij}-3\hat{{\bf r}}_{i}
\hat{{\bf r}}_{j}\big) \big( \cos q r +qr \sin q r \big) &  \nonumber \\
&  -\big(\delta_{ij}-\hat{{\bf r}}_{i}\hat{{\bf r}}_{j}\big)q^{2}r^{2}
\cos qr \Big] & \label{eq:retarded-dip-int}
\end{eqnarray}
where $\hat{{\bf r}}_{i}=r_{i}/r$.
For a \emph{fixed orientation} of the interatomic axis with respect to the
external field, (\ref{eq:tpot}) and (\ref{eq:retarded-dip-int}) give the 
well known $r^{-3}$ variation of the interaction energy at near-zone 
separations ($qr \ll 1$). 
The near zone limit of $U({\bf r})$ is \emph{strongly anisotropic}. 
It was noted by Thirunamachandran \cite{thirunamachandran80}
that when an \emph{average over all orientations} of the interatomic
axis with respect to the incident radiation direction is taken,
the static dipolar part of the coupling (i.e., the instantaneous,
non-retarded part  $r^{-3}\big(\delta_{ij}-3\hat{{\bf r}}_{i}
\hat{{\bf r}}_{j}\big)$) vanishes. 
The remaining `transverse' part is, in the near-zone, an attractive $r^{-1}$
potential. It is weaker by a factor of $(qr)^{2}$ than the $r^{-3}$
term.

However, thus far no scheme has been suggested wherein 
an average over all orientations is guaranteed for cold gases. 
We shall consider a spatial configuration of external fields which 
\emph{enforces the `averaging out'} of the $r^{-3}$ interactions.
A simple combination which ensures the suppression of the $r^{-3}$
interaction while retaining the weaker $r^{-1}$ attraction
in the near-zone, uses three orthogonal circularly polarized
laser beams pointing along $\hat{{\bf x}},\hat{{\bf y}},\hat{{\bf z}}$ 
(`a triad'---see Fig.\ 1).
Let us momentarily ignore interference between the three beams, and only
consider the sum of their intensities. 
In the near zone one can Taylor-expand Eqs.\ (\ref{eq:tpot}) 
and (\ref{eq:retarded-dip-int}) 
in powers of the small quantity $qr$.
Using the identity 
$\hat{{\bf e}}^{\ast (\pm)}_{i}({\bf q}) \hat{{\bf e}}^{(\pm)}_{j}
({\bf q})=\frac{1}{2}
\left( \left(\delta_{ij}-\hat{{\bf q}}_{i} \hat{{\bf q}}_{j} \right)
\pm {\mathrm{i}} \epsilon_{ijk}\hat{{\bf q}}_{k} \right)$, with $+(-)$
corresponding to left(right) circular polarizations,
together with Eqs.\ (\ref{eq:tpot}) and (\ref{eq:retarded-dip-int}),
the triad can be shown to give rise to the (near-zone) $r^{-1}$ 
pair potential
\begin{eqnarray}
\lefteqn{U({\bf r})  =  
-\frac{3 I q^{2} \alpha^{2}}{(16 \pi c \varepsilon_{0}^{2})} 
\times} & \quad & \quad \nonumber \\
 \quad  & \quad &  \frac{1}{r} \Big[\frac{7}{3} +
\left(\sin \theta \cos \phi \right)^{4} +
\left(\sin \theta \sin \phi \right)^{4} +
\left( \cos \theta \right)^{4} 
\Big]. 
\label{eq:triad}
\end{eqnarray}
Note that this interaction is attractive for any orientation
 $(\theta,\phi)$ of ${\bf r}$
relative to the beams as long as the polarizability $\alpha(q)$ is
real.
\begin{figure}[h]
\vspace{-1.6cm}
\begin{center}
\centerline{\psfig{figure=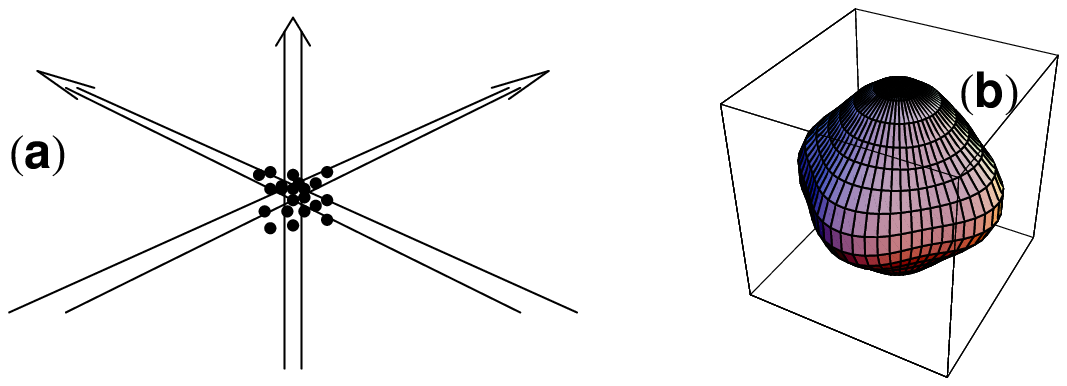,height=16cm}}
\end{center}
\vspace{-12.4cm}
\begin{caption}
{\textbf{(a):} Schematic depiction of a triad of lasers incident
upon an ensemble of atoms. 
This triad generates the attractive $r^{-1}$ potential given by Eq.\ (3), 
 whose magnitude has the angular dependence shown
in \textbf{(b)}.}
\end{caption}
\end{figure}

If one wishes, the angular anisotropy in (\ref{eq:triad}) can be 
cancelled to give a purely radial $r^{-1}$ potential
by combining a number of such triads with different orientations.
It is convenient to define the orientation of each triad by
the Euler angles $(\alpha,\beta,\gamma)$ 
\cite{mathews-walker}, namely, a rotation of $\alpha$ about
the $\hat{{\bf z}}$ axis, followed by a rotation of $\beta$ about
the new $\hat{{\bf y}}$ axis and finally a rotation of $\gamma$ about the
final $\hat{{\bf z}}$ axis.
One configuration that cancels the anisotropy completely
uses 6 triads (18 laser beams)
rotated through the following Euler angles:
$(0,\pi/4,\pi/8)$, $(0, \pi/4, -\pi/8)$, $(0, \pi/4, 3\pi/8)$,
$(0, \pi/4 ,-3 \pi/8)$, $(0,0,\pi/8)$, $(0,0, -\pi/8)$.
The last two triads should be of half the intensity $I$ of the others.
Then the interatomic potential becomes
\begin{equation}
U({\bf r})  =  -
\frac{11}{4 \pi}
\frac{I q^{2} \alpha^{2}}{c \varepsilon_{0}^{2}}
\frac{1}{r} = -\frac{u}{r}\;. 
\label{eq:supertriads}
\end{equation}

The main difficulty in realizing the near-zone $r^{-1}$ potential is that the
$r^{-3}$ interaction survives due to the
\textit{interference} between different pairs of beams, whose contribution is
proportional to the product of their respective field amplitudes.
This difficulty can be overcome if one introduces frequency shifts
between the laser beams.
Spreading the  frequencies $\omega_{n}$ of the $E_n$
laser fields ($n=1,2,3$ for one triad or $n=1,2 \dots 18$ for six triads) in
intervals about the central frequency makes the $r^{-3}$
interference terms
in the interaction energy $\propto E_n E^{\ast}_{n'}$, ($n\neq n'$),
oscillate at the difference frequencies $| \omega_{n}-\omega_{n'} |$.
If these difference frequencies are much higher than the other
relevant frequencies
(e.g., collective oscillation frequencies)
then the contribution of the interference terms
to the mean field potential averages to zero.
Typically these conditions hold for
 $| \omega_{n}-\omega_{n'} | \geq 10^4$ Hz.
Angular misalignment errors, $\delta$, between the orthogonal beams
should satisfy $\delta\ll q\, L$  
(where $L$ is the mean radius of the condensate)
and intensity fluctuations should satisfy
$\Delta I/I \ll q \,L$, in order to ensure the $r^{-3}$ cancellation
for the non-interfering terms $\propto \sum_{n} | E_{n} |^{2}$.
Although these oscillating $r^{-3}$ terms do not contribute
to the mean field potential, they can eject atoms from the condensate,
but this process can be strongly reduced, 
as will be discussed at the end of this letter.

A \emph{lower bound} on the magnitude of the
$r^{-1}$ attraction is obtained by using
the static rather than dynamic polarizability, which for sodium atoms,
 say, has the value $24.08 \times 10^{-24} {\mathrm{cm}}^{3}$. 
Thus, for a strongly off-resonant light intensity of
 $I = 10^{4}$ ${\mathrm{Watts/cm}}^{2}$ one finds
$-u/r$  $\approx$ $-2 \times 10^{-19} {\mathrm{eV}}$, at $r = 100$nm, 
the mean separation in a typical BEC.
This is only around $10^{-4}$  of the
magnitude of the van der Waals-London dispersion energy at this distance.
However in a system of many atoms the $r^{-1}$ potential acts 
over the entire sample whereas the van der Waals-London interaction is 
only effective for nearest-neighbors and so the $r^{-1}$ contribution
to the total energy can become important.

Our treatment of the many particle problem 
is based on a two-body potential 
$V({\bf r})=4 \pi a \hbar^{2} \delta({\bf r})/m - u/r$,
where the first term is the the pseudo-potential arising from the 
s-wave scattering ($a$ is the s-wave scattering length
and $m$ the atomic mass).
In order to write $V$ in this form we require that the $-u/r$ potential
be sufficiently weak (compared with the mean kinetic energy per particle)
so as not to affect the short-range hard-sphere scattering.
This requirement certainly holds if 
$a_* \gg \lambda_{DB} \gg a$, where 
$\lambda_{DB}$ is the de Broglie wavelength and
$a_*=h^2/m u$ is the Bohr radius associated with the 
gravitational-like coupling $u$.
With the values given above $a_*$  $\sim 10^3$ m, 
whilst for a typical BEC $\lambda_{DB} \sim
10^{-5}-10^{-3}$ m and 
$a\sim 3$ nm.

Consider now the application of this two-body potential
to a trapped dilute BEC gas well below the critical temperature. 
We assume that the condensate initially occupies
a fraction of the wavelength of the laser so that the near-zone
condition is valid (lasers in the far infrared, or microwave sources
would satisfy this condition).
The ``zero-temperature'' many-body problem leads, within
the MFA, to the following equation for
the order parameter $\Psi ({\bf R},t)$
\begin{equation}
i \hbar  \frac{\partial \Psi ({\bf R},t)}{\partial t} =
\left[
- \frac{\hbar^{2}}{2m} \nabla^{2}
+ V_{\mathrm{ext}}({\bf R})
+ V_{\mathrm{H}}({\bf R})
\right] \Psi({\bf R},t)  \label{21}
\end{equation}
where $V_{\mathrm{ext}}(R)= m\omega_{0}^2 R^2/2$ is for simplicity
an isotropic trap potential
(which can be set to zero in certain cases---see below), and
 $V_{\mathrm{H}}({\bf R})$ is the self-consistent Hartree potential
\begin{equation}
V_{\mathrm{H}}({\bf R})=
\frac{4 \pi a \hbar^{2}}{m} \mid \Psi({\bf R},t) \mid^{2}
- u\int d^3 R^{\prime}
\frac{ \mid \Psi({\bf R^{\prime}},t) \mid^{2}}
{\mid {\bf R}^{\prime}- {\bf R}\mid} \;. \label{22}
\end{equation}
The order parameter $\Psi( {\bf R},t)$
is normalized so that $\int d^{3} R \,\,|\Psi ( {\bf R},t) |^2$
$=N$,
with $N$ the total number of atoms. 
The usual Gross-Pitaevskii (GP) equation \cite{dalfovo99} is 
recovered in the limit when $u=0$.
The MFA is valid when the system is dilute, i.e.\,
$\rho a^{3} \ll 1$, with $\rho$ the density. 
An additional condition on the MFA validity is that the
$r^{-1}$ potential is weak, and this can be expressed as $\rho a_*^3 \gg 1$.
This constraint, as in the related problem of the charged Boson gas
\cite{foldy61}, means that many atoms must be present within the 
interaction volume $a_*^3$.
However, it is in fact the diluteness condition that turns out to be
the stricter of the two, as one can check from the MFA
 results for the density (see Table 1 below).

An analytical estimate for the mean radius 
of the N-atom condensate
can be given using the following variational wave function
\begin{equation}
\Psi_{\lambda}(R)=N^{\frac 12}(\pi \lambda^2 l^2_{0})^{-{3\over4}}
\exp (-R^2/2\lambda^2 l^2_{0} )
\label{eq:ansatz}
\end{equation}
where $l_0=\sqrt{\hbar/m\omega_{0}}$.
The variational parameter  $\lambda$  is proportional to the root mean
square  radius through $\sqrt{\langle R^2 \rangle }
 = \sqrt{3/2}\ \lambda l_{0} $.
This parameter is obtained by minimizing
the variational mean field energy
\begin{equation}
\frac{H(\lambda)}{N\;\hbar\omega_0}=
\frac{3}{4}
\left(
\lambda^{-2}
+\lambda^{2}
- 2 \tilde{u} \,\lambda^{-1}
+ \frac 23 \tilde{s} \,\lambda^{-3}
\right)
\label{varene}
\end{equation}
where we have chosen the dimensionless $\tilde{u}$ (proportional to the 
``gravity'' strength $u$)
and $\tilde{s}$ (proportional to s-wave scattering length $a$) to be
\begin{eqnarray}
\tilde{u} &=& 
\pi\sqrt{32\pi/9}
\; \; (N l_{0} / a_{*}) \; 
\label{eq:eps}\nonumber \\
\tilde{s} &=& 
\sqrt{2/\pi}
\; \; (N a/l_{0}) \;.
\label{eq:eta}
\end{eqnarray}
The numerical factors 
are chosen to make the equation for $\lambda$ simple
\begin{equation}
- \lambda^{-4} + 1 + \tilde{u} \,\lambda^{-3}
- \tilde{s} \,\lambda^{-5} = 0   \;. \label{vareq}
\end{equation}
This equation is equivalent to requiring
that the variational solution satisfies the
following virial relation:
$- T + V_{\mathrm{ext}} - \frac{1}{2} U_{u}
- \frac{3}{2} U_{s}$ $= 0$,
where $T$, $V_{ext}$, $U_{u}$ and $U_{s}$ are  the kinetic energy,
the harmonic trap potential energy, and the internal energies due to the
$-u/r$ and hard-sphere interatomic potentials, respectively. 
This relation can be obtained from scaling considerations
(see Ref. \cite{dalfovo96} for the case $u=0$).

The general asymptotic properties of the ground state solutions
of Eqs.\ (\ref{21}) and (\ref{22}), 
as a function of  $(\tilde{u},\tilde{s})$,
are summarized in the ``phase diagram'' of Fig.\ 2a
for positive scattering lengths. 
In this diagram
there are four asymptotic regions:
The non-interacting ideal region
($I$) and the  ordinary Thomas-Fermi region ($TF-O$)
are dominated by the balance of
the external trap potential with, respectively,  the kinetic energy
and the repulsive s-wave scattering, and so
are not sensitive to the $-u/r$ potential.
The regions $G$ and $TF-G$, which represent two \emph{new physical regimes}
for atomic BECs, are controlled by the balance
of the gravity-like potential with either the kinetic energy ($G$) or
the s-wave scattering ($TF-G$).
Neither region is sensitive to the external trap, so that
we can adiabatically turn it off ($V_{\mathrm{ext}}=0$)
and access either the $G$ 
or the $TF-G$ region.
\begin{figure}[h]
\vspace{-2cm}
\begin{center}
\centerline{\psfig{figure=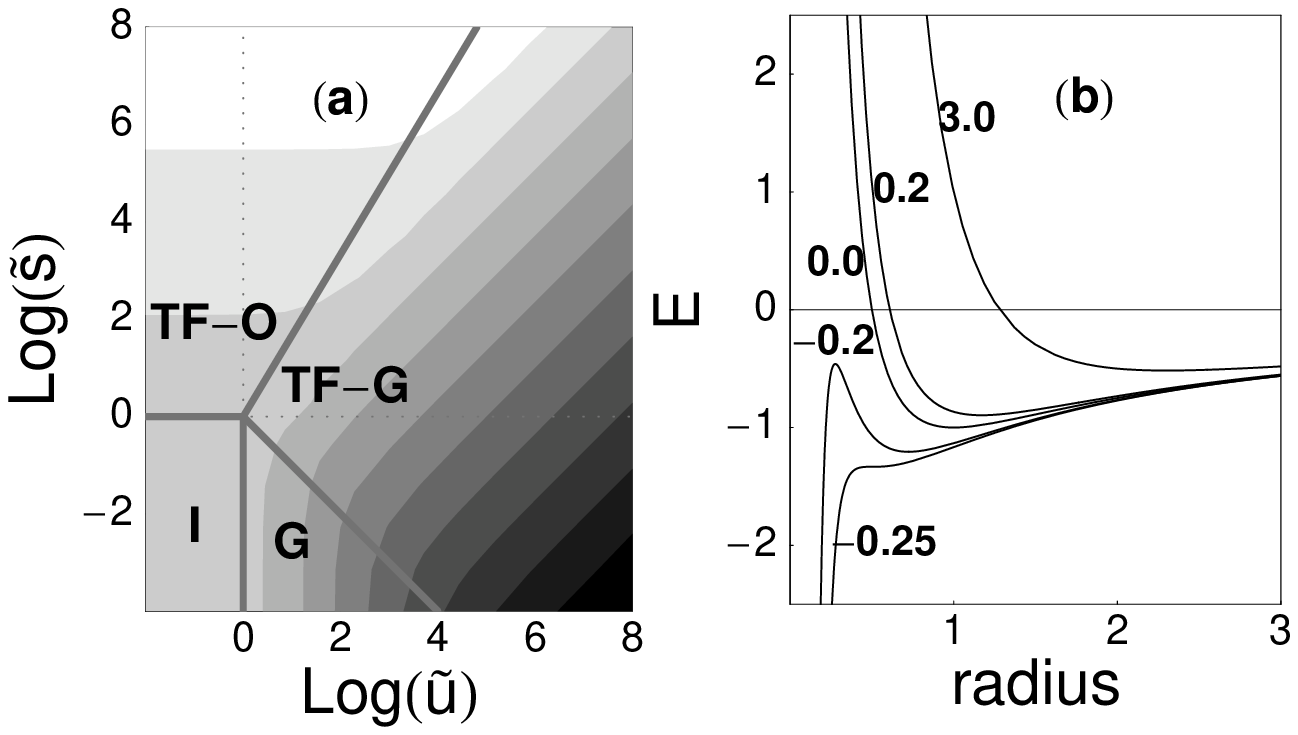,height=14cm}}
\end{center}
\vspace{-9cm}
\begin{caption}
{\textbf{(a)}:
Contour plot of  $\log (\lambda)$,
where $\lambda$ is the condensate radius,
in the parameter space $\log(\tilde{u})$ 
versus $\log(\tilde{s})$: 
darker shade corresponds to smaller $\lambda$.
The border separating the $TF-O$ and $TF-G$ regions is given by
$\tilde{s} = \tilde{u}^{5/3}$ 
and that separating the $TF-G$ and $G$ regions by
$\tilde{s} = \tilde{u}^{-1}$. 
\textbf{(b)}:
Mean energies per particle for large $\tilde{u}$
(no external trap) as a function of the condensate radius.
Curves are plotted for positive as well as negative values
of the scattering strength  $\tilde{s}/\tilde{u}$.
For $\tilde{s}/\tilde{u} \le -1/4$ there is no  minimum for 
a finite radius.
The energy and radii units are $\tilde{u}^2\hbar\omega_0$ and 
$l_0/\tilde{u}$, respectively.}
\end{caption}
\end{figure}
Experimentally, direct signatures of the $r^{-1}$ interaction
come from the radius $\lambda$ and the release energy $E_{rel} = T + U_{s}$.
The release energy is the kinetic energy that can be 
measured after the expansion occurring due to 
switching off the external trap and the laser fields \cite{dalfovo99}.
Table 1 summarizes these quantities 
as well as the peak density $\rho_{\mathrm{max}}$ in the four regions.
\begin{table}[h]
\begin{center}
\begin{tabular}{lllll} 
    & $G$ & $TF-G$ & $TF-O$ & $I$ \\ \cline{2-5} 
\raisebox{.0ex}[3.5ex]{defn.:} & 
 $\tilde{u} \gg 1$ & $\tilde{s} \ll \tilde{u}^{5/3}$
  & $\tilde{s} \gg 1$ &
 $\tilde{u} \ll 1$ \\ 
 &  $\tilde{s} \ll 1/\tilde{u}$  & $\tilde{s} \gg 1/\tilde{u}$ 
    & $\tilde{s} \gg \tilde{u}^{5/3}$ & $\tilde{s} \ll 1$ \\ \cline{2-5}
\raisebox{.0ex}[2.5ex]{$\lambda$:} & $1/\tilde{u}$ &
      $(\tilde{s}/\tilde{u})^{1/2}$
   & $\tilde{s}^{-1/5} $ & 1 \\ \cline{2-5}
\raisebox{.0ex}[3.5ex]{$E_{\mathrm{rel}}/ \hbar \omega_{0}$:} &
 $\frac{3}{4} \tilde{u}^{2}N$ 
    & $\frac{1}{2}  \tilde{u}^{3/2} \tilde{s}^{-1/2}N$
   & $\frac{1}{2} \tilde{s}^{2/5}N$
    & $\frac{3}{4}N$ \\ 
    & $\propto \; \; N^{3}$ & $ \propto \; \; N^{2} $
    & $ \propto \; \; N^{7/5}$ 
    & $\propto \; \; N $ \\ \cline{2-5}
\raisebox{.0ex}[3.5ex]{$\rho_{\mathrm{max}}:$} 
    & $\frac{\sqrt{32}^{3} \pi^{3} N^{4}}{27 a_{*}^{3}} $ 
    & $\frac{2 \pi^{2} N}{\sqrt{a a_{*}}^{3}}$ 
    & $\frac{15^{2/5} N^{2/5}}{8\pi a^{3/5} l_0^{12/5}}$
    & $ \frac{N}{\sqrt{\pi}^{3}l_{0}^{3}}$
\end{tabular}
\end{center}
\begin{caption}
{A comparison of the four asymptotic regions.
The radius, $\lambda$, and the release energy, $E_{\mathrm{rel}}$,
are discussed in the text. $\rho_{\mathrm{max}}$ is the peak density
(at the center) of the condensate.}
\end{caption}
\end{table}
We now focus on the properties of the two new regimes:
a) In the $TF-G$ region an analytic solution 
for the ground state of Eqs.\
(\ref{21}) and (\ref{22}) is given by
\begin{equation}
 \Psi_{TF-G} ({\bf R}) = \frac{\sqrt{N}}{2 R_{0}}
\sqrt{\frac{\sin(\pi R/R_{0})}{R}}
\Theta(R_{0} - R)   \; \label{eq:tf}
\end{equation}
where $R_{0}= \sqrt{a \ a_{*}}/2 $.
Contrary to the ordinary Thomas-Fermi limit of the GP equation,
 the size of the condensate is
fixed by the ratio of the coupling constants, $4 \pi a \hbar^{2}/m$
and $u$, and is
\textit{independent of $N$}. 
b)  
The $G$ region, where only the $r^{-1}$ attraction and 
kinetic energy play a r\^{o}le, is of particular interest
since our system is then equivalent to a Boson-star 
(a system of gravitating Bosons) \cite{ruffini69}
in the non-relativistic regime.
The mean field equations in this region are also identical to those
describing a single particle moving in the 
gravitational field generated by its own wavefunction \cite{penrose98}.
In both cases smooth bound solutions have been shown to exist 
\cite{ruffini69,penrose98}.
This establishes the possibility of a stable \textit{self-bound}
(no external trap) $r^{-1}$ condensate.

The gravitational-like attraction does not induce ``collapse'' of the 
condensate, since, at short radii, it is always weaker than the kinetic energy.
This can be seen from the scaling of the kinetic energy ($\lambda^{-2}$)  
versus that of the $r^{-1}$ potential ($\lambda^{-1}$)  
in Eq. (\ref{varene}) and Fig.\ 2b.
By contrast, this kind of instability can occur for \emph{negative} 
scattering lengths 
\cite{tiesinga92,goral99}
when $N$ exceeds a critical number ($N_{cr} \approx 0.6 \times |a|/l_0$)
because the mean energy due to scattering ($\lambda^{-3}$) is dominant
 at small radii.
The $u/r$ attraction does reduce, when combined with the attractive scattering,
the critical number  to $N_{cr}\approx 0.17 \times \sqrt{a_*/|a|}$
 (see Fig.\ 2b for the critical case with $\tilde{s}\tilde{u}=-1/4$).

Finally, we estimate the losses of $G$ or $TF-G$ condensates due to
the $r^{-3}$ oscillating \emph{interfering} terms discussed above.
Consider one of the possible oscillating interfering terms 
$A({\bf r}) \cos(\Omega t)$, where
$A(x,y,z) = - 3 \,u\, \frac{x\, y\,}{q^2\, r^5}$  
and $\Omega$ is the difference in frequency between the two 
interfering lasers.
Using Fermi's golden rule one can  derive an expression 
for the rate of depletion of the condensate density $|\Psi|^2$ 
due to creation of a pair of quasiparticles
of opposite momenta 
(with $k\approx \pm \sqrt{m\Omega / \hbar}$)
in the ideal homogeneous Bose gas 
\begin{equation}
\frac{d\,|\Psi|^2}{dt} = -\frac{\overline{|A(k)|^2}}{6\pi}
 |\Psi|^4 \left(\frac{m}{\hbar^2}\right)^{3/2}
\sqrt{\frac{\Omega}{\hbar}}
\label{eq.13}
\end{equation}
where  $\overline{|A(k)|^2}$ is the 
angular average of the square of  the Fourier transform 
of $A({\bf r})$ ($ 0.1418\, u^2/q^4 $).
For our purposes it is sufficient to apply Eq.\ (\ref{eq.13})
at each point ${\bf R}$ ($|\Psi|^2=|\Psi|^2({\bf R})$).
We then find the following approximations:
$d\, N_0/dt \approx \tilde{u}^{5}
\sqrt{\Omega\omega_0} /  (q l_0)^4$ in the $G$ region
and $d\, N_0/dt \approx \tilde{u}^{7/2} \tilde{s}^{-3/2}
\sqrt{\Omega\omega_0} / (q l_0)^4 $ in the $TF-G$ region.
These expressions can be used to find conditions
such that these loss rates are smaller than, say, the trap oscillation
frequency $\omega_0$. Taking, e.g.,
${\Omega} \approx 2\pi\times 10^4 $ s$^{-1}$, 
${\omega_0}\approx 2\pi\times 10^2$ s$^{-1}$, $ q l_0 \approx 1$,
$\tilde{u} \approx 5$, we obtain:
for  the $G$ region  
$d\;N_0/dt \approx 6 \times 10^{4}{\omega_0}$, i.e. we need more than
$10^5$ atoms;
for  the $TF$ region  (e.g., $ \tilde{s} \approx 1$),  
$d\;N_0/dt \approx 6 \times 10^{3}{\omega_0}$, i.e. we need more than
$10^4$ atoms.

To conclude, the laser-induced attractive $r^{-1}$ interaction
can give rise to stable condensates with unique static properties.
Their stability, long lifetime
(low loss rates incurred by the $r^{-3}$ oscillating terms) and
lack of sensitivity to alignment errors or amplitude noise of
the laser beams makes the experimental realization of such
condensates rather likely.
Their fascinating analogy with gravitating quantum systems whose gravitational
interaction can be enormously enhanced by the field merits further
research.

Support is acknowledged from ISF, Minerva and BSF (G.K.,S.G.),
the Royal Society (D.O.) and Arc-en-Ciel (V.A.,G.K.).
Valuable discussions with G.Hose are acknowledged by G.K..

\end{document}